# EFFECT OF SURFACE MICRO- AND NANOLAYERS OF COMPOSITE PARTICLES ON THERMOELECTRIC FIGURE OF MERIT (PHENOMENOLOGICAL APPROACH).

## A.A. Snarskii[1], M.I. Zhenirovskii[2]

[1] National Technical University of Ukraine "Kyiv Polytechnic Institute", Kyiv, Ukraine;
[2] N.N. Bogolyubov Institute of Theoretical Physics, Kyiv, Ukraine

- *Non-symmetric variants of self-consistent field theory are proposed that served the basis for theoretical investigation of the effect of thermal conductivity of composite particles surface layers on the effective thermoelectric figure of merit. It is shown that in some cases thermoelectric figure of merit can be essentially improved due to the presence of a shell.*

**Introduction**

Thermoelectric properties of composites are in the focus of constant attention (see, for instance, [1]). This attention is mainly due to the hope for obtaining material with a high thermoelectric figure of merit. These hopes are based on the exhibition of the unusual for conventional material values of local kinetic coefficients in micro- and nanolayers (see, for instance, [2]).

Below we shall consider composites of particles of relatively large (macroscopic) size with "standard" values of local kinetic coefficients. The unusual properties in this case will be inherent in the surface layer (shell, crystalline interface, etc.) of such particles. Thus, for instance, Ref. [2] reported on the dimensional effect in the electric conductivity and thermal conductivity of a thin layer between the composite particles.

One of the most adequate methods for calculation of effective properties of macroscopically inhomogeneous media is the theory of self-consistent field [3–5]. Such names as mean field theory, mean field approximation are also used; later on we shall use the acronym EMA (Effective Medium Approximation). A review on the use of EMA for calculation of average kinetic coefficients in composites with thermoelectric phenomena is presented, for instance, in [6]. In some cases other approximate methods are used as well: the Maxwell-Garnett and Landau-Lifshits approximations. For the problem under consideration (particle with a shell) all these approximations have one essential disadvantage. Let us explain it by example of two-phase conductive medium, and for simplicity thermoelectric effects will be assumed missing. Then, locally, Ohm's law will occur

$$\mathbf{j}(\mathbf{r}) = \sigma(\mathbf{r})\mathbf{E}(\mathbf{r}), \qquad (1)$$

where $\mathbf{j}(\mathbf{r})$ is electric current density, $\mathbf{E}(\mathbf{r})$ is field intensity, $\sigma(\mathbf{r})$ is electric conductivity taking on the values $\sigma_1$ and $\sigma_2 < \sigma_1$ in the first and second phases, respectively.

According to EMA, the equation for effective electric conductivity $\sigma_e$ which by definition interconnects average–volume fields $\langle \mathbf{E}(\mathbf{r}) \rangle$ and currents $\langle \mathbf{j}(\mathbf{r}) \rangle$ $\langle \mathbf{j}(\mathbf{r}) \rangle = \sigma_e(\mathbf{r})\langle \mathbf{E}(\mathbf{r}) \rangle$ is of the form

$$\frac{\sigma_e - \sigma_1}{2\sigma_e + \sigma_1}p + \frac{\sigma_e - \sigma_2}{2\sigma_e + \sigma_2}(1-p) = 0, \qquad (2)$$

where $p$ is first phase concentration.

Conclusion (2) is based on the calculation of a local field inside a spherical particle from the first or second phase immersed in a medium with the sought-for effective conductivity. This is the problem of mathematical physics on the so-called secluded particle with a homogeneous field on the infinity. However, even with a large concentration of particles of one phase in another, when the concept of "seclusion" is invalid, the concentration dependence of $\sigma_e$ in many cases describes well the experimental data.

One should note: approximation (2) means that with $p \ll 1$ a medium represents randomly scattered first-phase balls in the second-phase matrix, and with $1 - p \ll 1$ on the contrary – second-phase balls in the first-phase matrix. Thus, simultaneous replacement $\sigma_1 \longleftrightarrow \sigma_2$ and $p \longleftrightarrow 1 - p$ leaves $\sigma_e$ unchanged

$$\sigma_e(\sigma_1, \sigma_2, p) = \sigma_e(\sigma_2, \sigma_1, 1 - p). \tag{3}$$

In particular, it means that phases with conductivities $\sigma_1$ and $\sigma_2$ are "parity" ones, each of them with a small concentration representing a set of secluded inclusions. At the same time, phases in such structures as a mixture of particles (first phase) with shells (second phase) occupy quite different geometric positions, and relationship (3) should not occur. Thus, standard, "phase-parity" EMA has to be modified.

In this paper, two EMA modifications will be constructed, and on their basis the effect of second ("intercalation", "shell") phase on thermoelectric properties, including thermoelectric figure of merit, will be studied.

## 1. First EMA modification

In order to take into account the specific location of the second phase with conductivity $\sigma_2$, let us imagine inhomogeneous medium as a structure obtained by double consecutive use of EMA procedure. Consider first a case when thermoelectric effects are missing. The concentration of the well-conducting phase ($\sigma_1 \gg \sigma_2$) will be selected a little below the percolation threshold which within EMA in the three-dimensional case is equal to $p_c = 1/3$. Then at the first stage for effective conductivity $\sigma_{e1}$ we have

$$\frac{\sigma_{e1} - \sigma_1}{2\sigma_{e1} + \sigma_1} p_1 + \frac{\sigma_{e1} - \sigma_2}{2\sigma_{e1} + \sigma_2}(1 - p_1) = 0, \tag{4}$$

where $p_1$ is concentration of the well-conducting phase in a composite with effective conductivity $\sigma_{e1}$, $p_1 \leq p_c$.

Let us now consider that later on the role of the second phase will be played by homogeneous medium with conductivity $\sigma_{e1}$. The second construction stage is similar to the first one. The role of the first phase is still played by the phase with conductivity $\sigma_1$, but that of the second phase – by the phase with conductivity $\sigma_{e1}$, the concentration of which is selected equal to $1 - p_1$. The effective conductivity of such a medium is found from equation

$$\frac{\sigma_e - \sigma_1}{2\sigma_e + \sigma_1} p_1 + \frac{\sigma_e - \sigma_{e1}}{2\sigma_e + \sigma_{e1}}(1 - p_1) = 0. \tag{5}$$

Thus, full concentration of the first phase $\sigma_1$ is

$$p = p_1 + (1 - p_1)p_1. \tag{6}$$

Such construction of a medium with $\sigma_e$ from (5) implies that at the first stage (4) the first phase is surrounded by the second ($p_1 \leq p_c$). Current flowing through the medium must necessarily pass through the poorly conducting ($\sigma_2$) phase, and the medium on the whole is poorly conducting. At the second stage, $\sigma_{e1}$ plays the role of the poorly conducting phase, its concentration $1 - p_1$ being a little larger than the percolation threshold. Thus, at the second stage, too, inclusions of the well conducting phase are separated by the poorly conducting ($\sigma_{e1}$). Fig. 1 shows concentration dependences of the effective conductivity within the standard EMA (2) and according to modified (5). As can be seen from Fig. 1, for $\sigma_e$ (5) relationship (3) does not occur, while for standard EMA (2) in relationship $\sigma_e(\sigma_1,\sigma_2,p)$ and $\sigma_e(\sigma_2,\sigma_1,1-p)$ coincide.

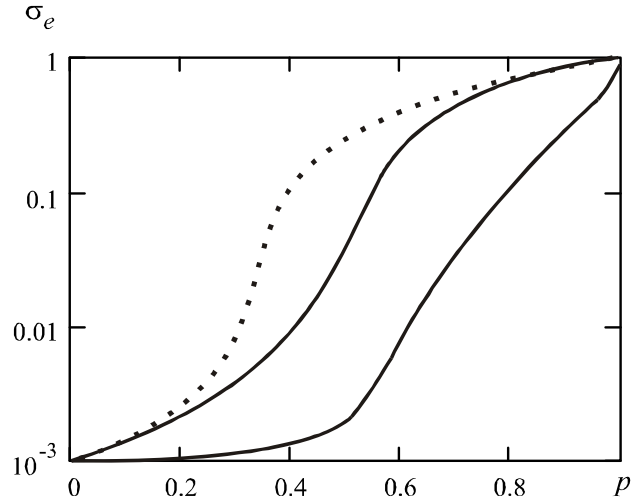

Fig. 1. Concentration dependences of effective conductivity $\sigma_e$ The upper plot – $\sigma_e(\sigma_1,\sigma_2,p)$ – represents two identically coincident dependences of standard EMA approximation $\sigma_{st}(\sigma_1,\sigma_2,p) \equiv \sigma_{st}(\sigma_2,\sigma_1,1-p)$, middle plot – $\sigma_e(\sigma_1,\sigma_2,p)$ solution of equation (5), lower plot – $\sigma_e(\sigma_2,\sigma_1,1-p)$. As an example, $\sigma_1 = 1$, $\sigma_2 = 10^{-3}$ conventional units were selected.

Prior to describing thermoelectric properties of structures described by the first EMA modification, let us consider another modification.

## 2. Second EMA modification

In this modification a medium under consideration consists of spherical particles of the first phase with the shell of the second phase immersed in a medium of the first phase. Thus, with a concentration of shelled particles a little larger than $1-p_1$, current necessarily has to pass through the poorly conducting phase shell.

Like in standard EMA, one should first solve the problem of a secluded particle, but this time the particle has a shell, see Fig. 2.

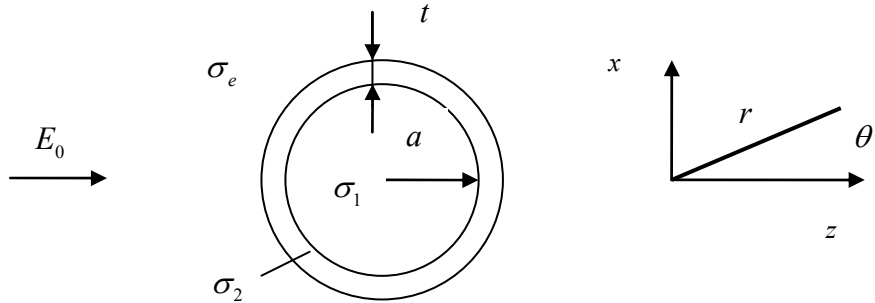

Fig. 2. Secluded shelled spherical particle. Homogeneous field $\mathbf{E}_0$ is assigned on the infinity. According to EMA schematic, shelled particle is immersed in a medium with the sought for effective conductivity $\sigma_e$.

The expression for electric field intensities inside a particle and in a shell can be found solving the problem of mathematical physics, on the assumption that for the electric field intensity **E(r)** and the electric field density **j(r)** the equations $\text{rot}\mathbf{E}(\mathbf{r}) = 0$ and $\text{div}\mathbf{j}(\mathbf{r}) = 0$ are met, and at the phase boundaries normal current density components and tangential electric field intensities are continuous.

Solution of this problem is well known, for example, to an accuracy of notations, it is given in [7]:

$$E_{1z} = A_1, \quad E_{2z} = A_2 - 2\frac{B_2}{r^3} + 3\frac{B_2}{r^3}\sin^2(\theta), \quad E_{2x} = -3\frac{B_2}{r^3}\sin(\theta)\cos(\theta)\cos(\varphi), \quad (7)$$

$$E_{1x} = E_{1y} = 0, \quad E_{2y} = E_{2x},$$

where $\varphi$ and $\theta$ are angles in a spherical coordinate system, and

$$A_1 = 9\frac{\sigma_e \sigma_2 b^3}{\Omega}E_0, \quad A_2 = 3\frac{\sigma_e(\sigma_1 + 2\sigma_2)b^3}{\Omega}E_0, \quad B_2 = 3\frac{\sigma_e(\sigma_2 - \sigma_1)a^3 b^3}{\Omega}E_0, \quad (8)$$

and $\Omega = \sigma_1\sigma_2(2a^3 + b^3) + 2(\sigma_e\sigma_1 + \sigma_2^2)(b^3 - a^3) + 2(2b^3 + a^3)$, $b = a + t$.

From (7) it is evident that a shelled field ($E_{2x}$ and $E_{2z}$) is coordinate dependent. At the same time, the use of EMA self-consistency pattern, when field distortions introduced by different phase inclusions, compensate each other, requires constancy of fields inside the inclusions. Therefore, another approximation is further needed. Here due to the fact that the shell will be considered thin ($t \ll a$), instead of $E_{2x}$ and $E_{2z}$ one can (this is the additional approximation) take their volume-averaged values $\langle E_{2x} \rangle$ and $\langle E_{2z} \rangle$. Under average $\langle ... \rangle$ we shall understand

$$\langle f \rangle = \frac{3}{4\pi\left[(a+t)^3 - a^3\right]}\int_0^\pi d\theta \int_0^{2\pi} d\varphi \int_a^{a+t} f(r,\varphi,\theta) \cdot r^2 \sin(\theta)dr. \quad (9)$$

Then for the shelled fields we shall get

$$\langle E_{2z} \rangle = A_2, \quad \langle E_{2x} \rangle = 0. \quad (10)$$

Let us now represent a particle with a shell as a single (new) phase, the average field $E_t$ in this phase particle being equal to

$$E_{tz} = \frac{V_1}{V}E_{1z} + \frac{V_2}{V}\langle E_{2z} \rangle, \quad V = V_1 + V_2, \quad (11)$$

where $V_1 = \frac{4}{3}\pi a^3$ is volume of particle from the first phase, and $V_2 = \frac{4}{3}\pi\left[(a+t)^3 - a^3\right]$ is shell volume.

Substituting $E_{1z}$ (7) and $\langle E_{2z} \rangle$ (10) into (11), we find

$$E_{tz} = \frac{a^3 A_1 + \left[(a+t)^3 - a^3\right]A_2}{b^3}, \quad (12)$$

which, just like (10), takes into account components with small parameter $t/a$ not larger than the first order.

Substituting into (12) the expression for $A_1$, $A_2$ and $B_2$ (8) in the first approximation of $t/a$ infinitesimal, we get

$$E_{tz} = \frac{3\sigma_e\left[\sigma_1 + (\sigma_1 + 2\sigma_2)\dfrac{t}{a}\right]}{\Omega_1 + \sigma_e\Omega_2}E_0, \quad (13)$$

where

$$\Omega_1 = \sigma_1\sigma_2\left(1+\frac{t}{a}\right)+2\sigma_2^2\frac{t}{a}, \quad \Omega_1 = 2\left[\sigma_1\frac{t}{a}+\sigma_2\left(1+2\frac{t}{a}\right)\right]. \tag{14}$$

According to EMA, now it is necessary to find a field inside the particles with conductivity $\sigma_1$, immersed in a matrix with conductivity $\sigma_e$, this field (see, for example, [8]) is equal to

$$E = \frac{3\sigma_e}{2\sigma_e+\sigma_1}E_0. \tag{15}$$

Self-consistency equation in this case is of the form
$$E_{tz}p_1 + E(1-p_1) = E_0. \tag{16}$$
where $E_{tz}$ (13) is an averaged field in a particle with a shell, the concentration of such particles $p_1$, $E$ (15) is a field in homogeneous particles consisting of phase with conductivity $\sigma_1$, their concentration $1-p_1$.

Substituting into (16) $E_{tz}$ (13), $E$ (15) and (14) we get a quadratic equation the solution of which can be written as

$$\sigma_e = \frac{-D_2+\sqrt{D_2^2-4D_1\sigma_1\Omega_1}}{2D_1}, \tag{17}$$

where

$$D_1 = 2Mp_1+3\Omega_2(1-p_1)-2\Omega_2, \quad D_2 = Mp_1\sigma_1+3\Omega_1(1-p_1)-2\Omega_1-\sigma_1\Omega_2,$$
$$M = 3\left[\sigma_2+(\sigma_1+2\sigma_2)\frac{t}{a}\right]. \tag{18}$$

Fig.3 shows concentration dependence of effective conductivity $\sigma_e$ according to (17).

Note that, as it must, a drastic change in the concentration behaviour of effective conductivity for given modification occurs not on a standard percolation threshold, which for EMA is equal to $1/3$, but on its symmetric one, equal to $2/3$, which in the standard case does not become apparent (see [9], though). With the concentration of shelled particles equal to $2/3$, they fully overlap current pass along conductivity phase, thus, current should necessarily pass through shelled particles, i.e. through the shell itself that has low conductivity.

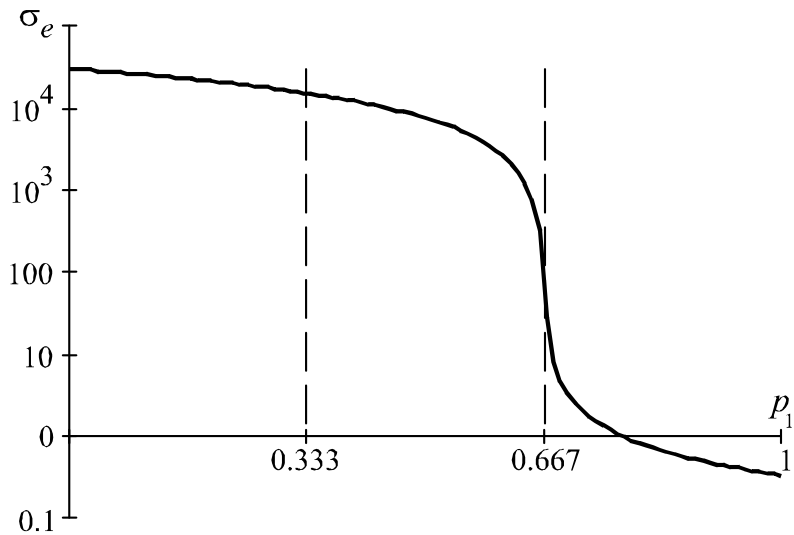

*Fig.3. Concentration dependence of effective conductivity $\sigma_e$ according to (17),*

$\sigma_1 = 10^5$, $\sigma_2 = 1$, *conventional units, selected as an example.*

Effective conductivity strongly depends on the intercalation thickness, Fig.4 shows this dependence for a concentration of shelled particles a little larger than $2/3$.

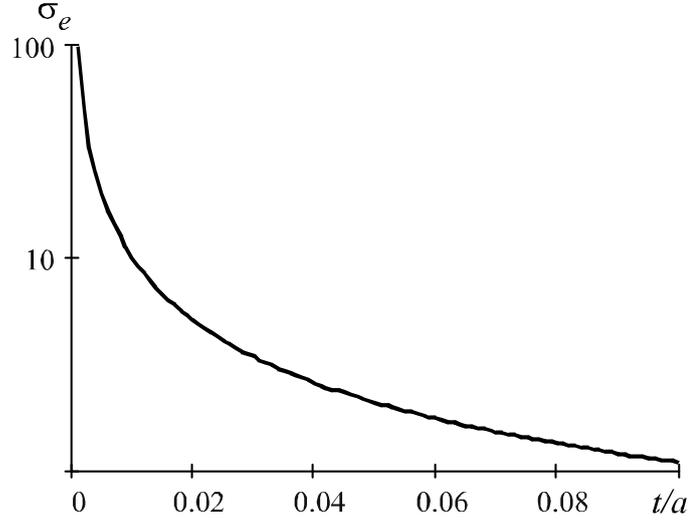

*Fig.4. Dependence of effective conductivity $\sigma_e$ on the intercalation thickness in the framework of the second modification (17). Concentration of shelled particles $p_1=2/3+0.1$.*

## 3. Thermoelectric properties

Here the following notations will be used:

$$\begin{cases} \mathbf{j} = \sigma \mathbf{E} + \sigma\alpha(-\nabla T), \\ \dfrac{\mathbf{q}}{T} = \sigma\alpha \mathbf{E} + \kappa(1+ZT)(-\nabla T), \quad ZT = \dfrac{\sigma\alpha^2}{\kappa}T, \end{cases} \quad (19)$$

where $\mathbf{j}$ is electric flux density, $\mathbf{q}$ is heat flow density created by two thermodynamic forces: $\mathbf{E}$ – electric field intensity ad $\nabla T$ – temperature gradient, $\alpha$ – thermoelectric coefficient, $\kappa$ – thermal conductivity, $Z$ – figure of merit (the Ioffe number).

In case of two-phase media, when in the absence of thermoelectric effects a concentration dependence of effective conductivity is known, the isomorphism method [10–12] allows "automatically" obtaining the concentration dependences of effective coefficients with regard to thermoelectric effects. In the most general form the isomorphism method was elaborated in [12] and realized not only for thermoelectric effects, but also for two- and three-dimensional galvanomagnetic effects and the problem of conductivity of anisotropic media.

Omitting the details of rather cumbersome calculations, we shall give only the final result [12]. Suppose we know the solution of a problem on calculation of effective electric conductivity $\sigma_e$ in a medium without thermoelectric effects which will be written as

$$\sigma_e = \sigma_1 f(p,h), \quad (20)$$

where $h = \sigma_2/\sigma_1$ and $p$ – as usual, is a concentration of the first (well conducting phase), and matrix of kinetic coefficients $\hat{\sigma}$ in thermoelectric case is of the form

$$\hat{\sigma}_i = \begin{pmatrix} \sigma_i & \gamma_i \\ \gamma_i & \chi_i \end{pmatrix}, \; \gamma_i = \sigma_i \alpha_i, \; \chi_i = \kappa/T + \sigma_i \alpha_i^2, \quad (21)$$

where $i = 1,2$ denotes phase number.

Then the effective kinetic coefficients are of the form

$$\begin{cases} \sigma_e = \dfrac{(\mu\sigma_1 - \sigma_2)f(p,\lambda) - (\lambda\sigma_1 - \sigma_2)f(p,\mu)}{\mu - \lambda}, \\[2mm] \alpha_e = \dfrac{(\mu\alpha_1\sigma_1 - \alpha_2\sigma_2)f(p,\lambda) - (\lambda\alpha_1\sigma_1 - \alpha_2\sigma_2)f(p,\mu)}{(\mu\sigma_1 - \sigma_2)f(p,\lambda) - (\lambda\sigma_1 - \sigma_2)f(p,\mu)}, \\[2mm] \kappa_e = \dfrac{\kappa_1\sigma_1(\mu - \lambda)f(p,\lambda)f(p,\mu)}{(\mu\sigma_1 - \sigma_2)f(p,\lambda) - (\lambda\sigma_1 - \sigma_2)f(p,\mu)}, \end{cases} \quad (22)$$

where functions $f(p,\lambda)$ and $f(p,\mu)$ are the same as in (20) (with $h$ substituted by $\lambda$ and $\mu$ respectively), and $\mu$ and $\lambda$ are described by expressions:

$$\begin{cases} \mu = \dfrac{1}{4\sigma_1\kappa_1}\left(\sqrt{\left(\sqrt{\sigma_1\kappa_2} + \sqrt{\sigma_2\kappa_1}\right)^2 + \sigma_1\sigma_2 T(\alpha_1 - \alpha_2)^2} + \sqrt{\left(\sqrt{\sigma_1\kappa_2} - \sqrt{\sigma_2\kappa_1}\right)^2 + \sigma_1\sigma_2 T(\alpha_1 - \alpha_2)^2}\right), \\[3mm] \lambda = \dfrac{1}{4\sigma_1\kappa_1}\left(\sqrt{\left(\sqrt{\sigma_1\kappa_2} + \sqrt{\sigma_2\kappa_1}\right)^2 + \sigma_1\sigma_2 T(\alpha_1 - \alpha_2)^2} - \sqrt{\left(\sqrt{\sigma_1\kappa_2} - \sqrt{\sigma_2\kappa_1}\right)^2 + \sigma_1\sigma_2 T(\alpha_1 - \alpha_2)^2}\right). \end{cases} \quad (23)$$

Let us now consider the thermoelectric properties of composite in the framework of the first modification. Choosing function $\sigma_e = \sigma_1 f(p,h)$ from (20) according to solution (5) and substituting it into (22), we find the effective kinetic coefficients and thermoelectric figure of merit. Fig.5 shows thermoelectric figure of merit as a function of second phase thermal conductivity.

One should note the nonmonotonous behaviour of this dependence. First, with reduction of second phase thermal conductivity, the effective thermal conductivity is reduced as well, leading to increase in the effective figure of merit. However, with subsequent reduction of thermal conductivity, the figure of merit starts dropping. This drop is due to the fact that a change in second phase thermal conductivity changes not only the effective thermal conductivity of composite, but also the effective electric conductivity and thermoEMF.

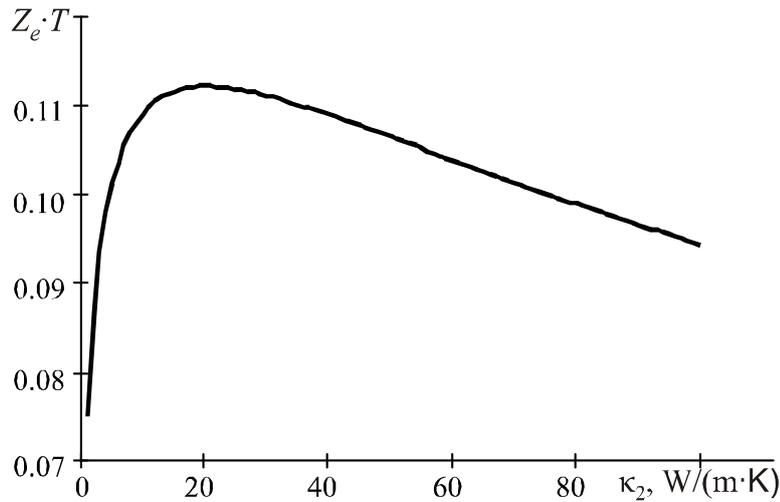

*Fig.5. Dependence of thermoelectric figure of merit of composite described by the first EMA modification on the second phase thermal conductivity. As an example, the values of local kinetic coefficients*

*of the first and second phases were selected as follows: $\sigma_1 = 10^7$ Ohm$^{-1}$m$^{-1}$, $\sigma_2 = 0.3 \cdot 10^7$ Ohm$^{-1}$m$^{-1}$, $\kappa_1 = 100$ W/m·K, $\alpha_1 = 10^{-4}$ V/K, $\alpha_2 = 10^{-10}$ V/K, $T = 300$ K.*

Let us now consider the thermoelectric properties of composite described by the second EMA modification. To do this, we shall use the same isomorphism method, where now as a function $f$ from (20) we shall take $\sigma_e / \sigma_1$ from (17). Fig.6 shows dependences of composite figure of merit on the intercalation thickness. As can be seen from the figure, the effective figure of merit grows with increasing intercalation thickness.

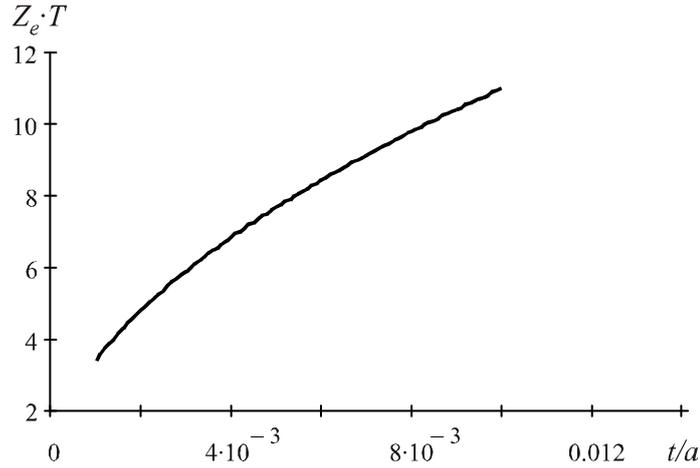

*Fig.6. Dependences of effective figure of merit on the intercalation thickness.*
*As an example, the values of local kinetic coefficients of the first and second phases were selected as follows: $\sigma_1 = 10^7$ Ohm$^{-1}$m$^{-1}$, $\sigma_2 = 0.3 \cdot 10^7$ Ohm$^{-1}$m$^{-1}$, $\kappa_1 = 100$ W/m·K, $\alpha_1 = 2 \cdot 10^{-4}$ V/K, $\alpha_2 = 1.9 \cdot 10^{-4}$ V/K, $T = 300K$.*
*Concentration of shelled phase is equal to $2/3 + 10^{-2}$, i.e. exceeds a little the second percolation threshold. The relative percolation thickness $t/a$ (see, Fig.2) changes within $0.001 \div 0.012$.*

Fig.7 shows dependences of effective figure of merit on the intercalation thermal conductivity. Three cases of different values of intercalation thickness are considered, their relative values being selected equal to: $10^{-2.8}$ – upper plot, $10^{-3}$ – medium plot and $10^{-3.2}$ – lower plot. As can be seen from the figure, the largest effective figure of merit growth with thermal conductivity reduction occurs in the thickest intercalation.

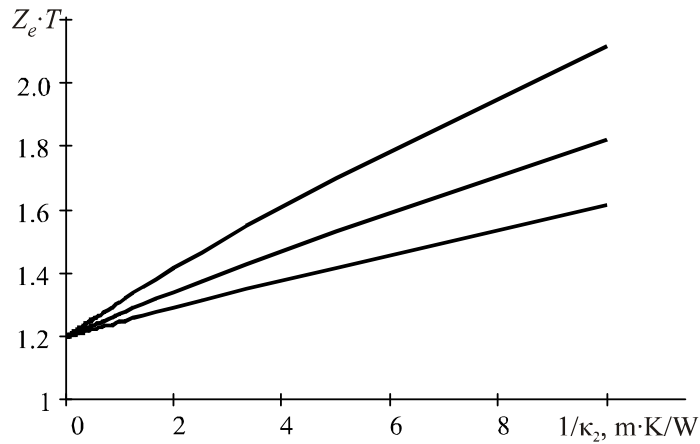

*Fig.7. Dependences of effective figure of merit on the intercalation thermal conductivity.*
*As an example, the values of local kinetic coefficients of the first and second phases were selected as follows: $\sigma_1 = 10^7$ Ohm$^{-1}$m$^{-1}$, $\sigma_2 = 0.3 \cdot 10^7$ Ohm$^{-1}$m$^{-1}$, $\kappa_1 = 100$ W/m·K, $\alpha_1 = 2 \cdot 10^{-4}$ V/K, $\alpha_2 = 1.9 \cdot 10^{-4}$ V/K, $T = 300K$.*
*Concentration of shelled phase is equal to $2/3 + 10^{-2}$, i.e. exceeds a little the second percolation threshold.*

*The relative intercalation thickness $t/a$ is equal to $10^{-2.8}$ – upper plot,*
*$10^{-3}$ – middle plot and $10^{-3.2}$ – lower plot.*

**Conclusions**

Investigation of thermoelectric properties of macroscopically inhomogeneous composite shows the essential role of intercalation (layer) between separate phases. Both composite models obtained by modification of medium field theory result in the same basically important qualitative conclusion – increase in efficient thermoelectric figure of merit with reduction of intercalation thermal conductivity (even with its infinitesimal volume with respect to the entire composite bulk). At the same time, the resulting quantitative values for these models are different. Further progress in this direction seems possible only in close contact with the experiment.